\documentclass{svjour3}                     % onecolumn (standard format)
\smartqed  % flush right qed marks, e.g. at end of proof
\usepackage{graphicx}
\newcommand{\beq}{\begin{equation}}
\newcommand{\eeq}[1]{\label{#1} \end{equation}} 

\begin{document}

\title{$K^-$ absorption in nuclei by two and three nucleons%\thanks{Grants or other notes
}


\author{V.K.~Magas \and
E.~Oset \and
A.~Ramos
}


\institute{V.K.~Magas \at
              Departament d'Estructura i Constituents de la Mat\'eria,\\
              Universitat de Barcelona,  Diagonal 647, 08028 Barcelona, Spain \\
              Tel.: +34-93-4039188\\
              Fax: +34-93-4021198\\
              \email{vladimir@ecm.ub.es}           %  \\
           \and
           E.~Oset \at
              Departamento de F\'{\i}sica Te\'orica and \\
              IFIC Centro Mixto Universidad de Valencia-CSIC\\
              Institutos de Investigaci\'on de Paterna,\\
               Apdo. correos 22085, 46071, Valencia, Spain
					\and
           A.~Ramos \at
              Departament d'Estructura i Constituents de la Mat\`eria\\
              Universitat de Barcelona,  Diagonal 647, 08028 Barcelona, Spain 
}

\date{Received: date / Accepted: date}

\maketitle

\begin{abstract}
It will be shown that the peaks in the ($\Lambda p$) and ($\Lambda d$) invariant mass distributions, observed in recent 
FINUDA experiments and claimed to be signals of deeply bound kaonic states, 
are naturally explained in terms of $K^-$ absorption by two or three nucleons leaving the rest of the original nuclei as spectator. 
For reactions on heavy nuclei, the subsequent interactions of the particles produced in the primary absorption process  with the residual nucleus play an important role. Our analyses leads to the conclusion that at present there is no experimental evidence of deeply bound $K^-$ state in nuclei. 
Although the FINUDA experiments have been done for reasons which are not supported a posteriori, some new physics can be extracted from the data. 

\end{abstract}

\section{What is the present experimental situation with $K^-$-nucleons bound states?}
\label{sec-1}

The possibility of having deeply bound $K^-$ states in nuclei is recently receiving much
attention both theoretically and experimentally. An overview of the different 
theoretical approaches and their results can be found, for example, in Ref.  \cite{Magas:2008bp}.

On the experimental side the situation is still at a very early stage. Initial hopes
that a peak seen in the  $(K^-_{stop},p)$ reaction  on $^4$He
\cite{Suzuki:2004ep} could be a signal of a $K^-$ bound in the trinucleon with
a binding of 195 MeV gra\-dua\-lly lost a support. First, an alternative
explanation of the peak was presented in \cite{toki}, 
showing that a peak with the strength claimed in the experiment
was coming from $K^-$ absorption on a pair of nucleons going to $p \Sigma$, 
leaving the other two nucleons
as spectators. This hypothesis led to the prediction that such a peak should
be seen in other light or medium nuclei where it should be narrower and weaker
as the nuclear size increases. This was
confirmed with the finding of such a peak in the $(K^-_{stop},p)$ reaction on
$^6$Li, which already fades away in $^{12}$C nuclei at FINUDA 
\cite{agnellonuc}. In \cite{toki} the $K^-$ absorption was described as taking place on  
 (np) pairs of the Fermi sea. In \cite{agnellonuc} the same explanation was
given for the peaks suggesting that
the (np) pairs would be correlated in "quasi"-deuteron clusters. 
The final development in this discussion
 has come from a new experiment of the KEK reaction of \cite{Suzuki:2004ep} 
 reported in 
 \cite{Sato:2007sb} where,
 performing a more precise measurement, which amended deficiencies in the efficiency
 corrections, 
 the relatively narrow
 peak seen in \cite{Suzuki:2004ep} disappears
 and only a  broad bump remains around the region where the peak was
 initially claimed. The position and width of this broad bump are in
 agreement with the estimations done in \cite{hyper,npangels} based on the 
 kaon absorption mechanism of \cite{toki}.

    The second source of initial hope came from the experiment of the FINUDA
    collaboration 
    \cite{Agnello:2005qj}, where a peak seen in the invariant mass distribution
    of $\Lambda p$ following $K^-$ absorption in a mixture of light nuclei was
    interpreted as evidence for a $K^- pp$ bound state, with 115 MeV binding and
    67 MeV width. 

\begin{figure}
%  \includegraphics[width=1.0\textwidth]{Gamma_3N_29_vs_3d.eps}
 \includegraphics[width=0.75\textwidth]{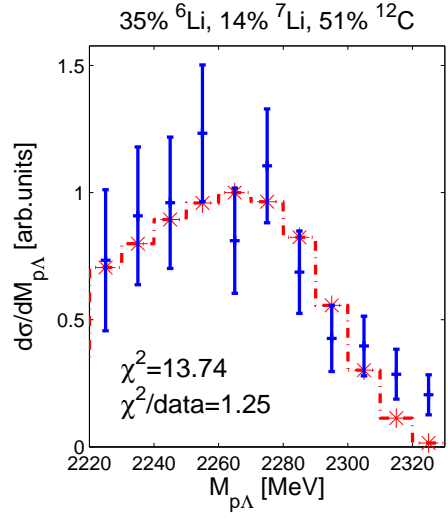}
\caption{Invariant mass of $\Lambda p$
distribution for $K^-$ absorption in light nuclei in the following proportion \cite{Agnello:2005qj}:  
51\% $^{12}$C,  35\% $^{6}$Li and 14\% $^{7}$Li, including
kaon boost machine corrections \cite{future_finuda}. Stars and histogram show 
result of our calculations \cite{Magas:2006fn}, experimental points and errorbars are taken from \cite{Agnello:2005qj}.
}
\label{fig2}       % Give a unique label
\end{figure}

    However, it was shown in \cite{Magas:2006fn,Kpp,Ramos:2007zz} that the peak
    seen is naturally explained in terms of $K^-$ absorption on a pair of nucleons
    leading to a $\Lambda p$ pair, followed by the rescattering of the $p$ or the
    $\Lambda$ on the remnant nucleus - see Fig. \ref{fig2}. 
    
    More recently, a new experiment of the FINUDA collaboration \cite{:2007ph}
    found a peak on the invariant mass of $\Lambda d$ following the absorption
    of a $K^-$ on $^6$Li, which was interpreted as a signature for a bound
    $\bar{K}NNN$ state with 58 MeV binding and 37
    MeV width. These results are puzzling, since 
    the bound state of the ${\bar K}$ in the three nucleon system has 
    significantly smaller values for the binding and width than those claimed
    for the bound state of the ${\bar K}$ in the two nucleon system 
    \cite{Agnello:2005qj}.
    
    About the same time as the FINUDA experiment \cite{:2007ph}
    a similar  experiment was performed at KEK 
    \cite{Suzuki:2007kn}, looking also at the $\Lambda d $ invariant mass 
    following $K ^-$ absorption but on a $^4$He target. The authors of this 
    latter work do not
    share the conclusions of  \cite{:2007ph} concerning the association of the
    peak to a $\bar{K}$ bound state, and claim instead that the peak could be
     a signature of three body absorption.

\begin{figure*}
%  \includegraphics[width=1.0\textwidth]{Gamma_3N_29_vs_3d.eps}
 \includegraphics[width=0.5\textwidth]{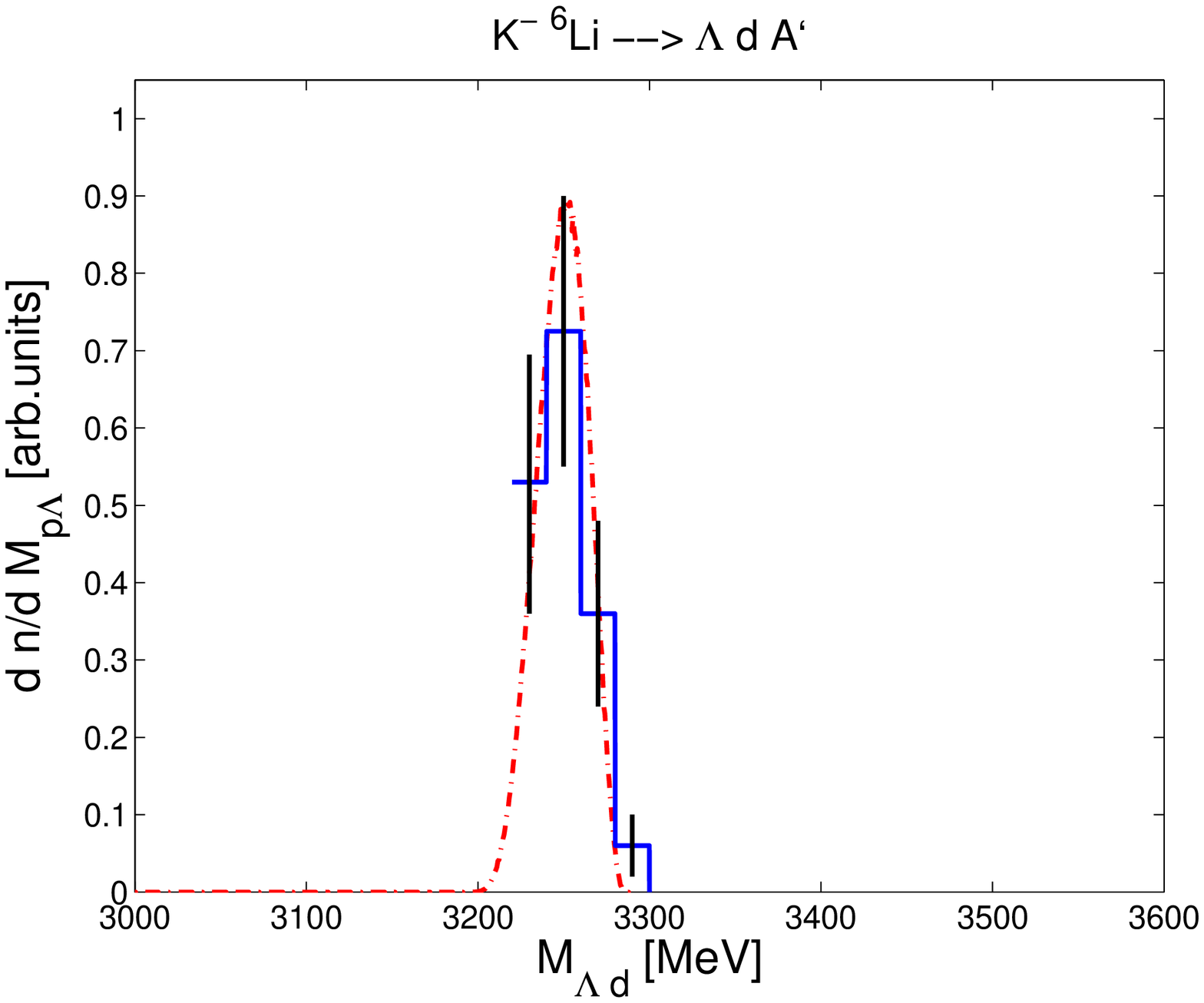}
 \includegraphics[width=0.5\textwidth]{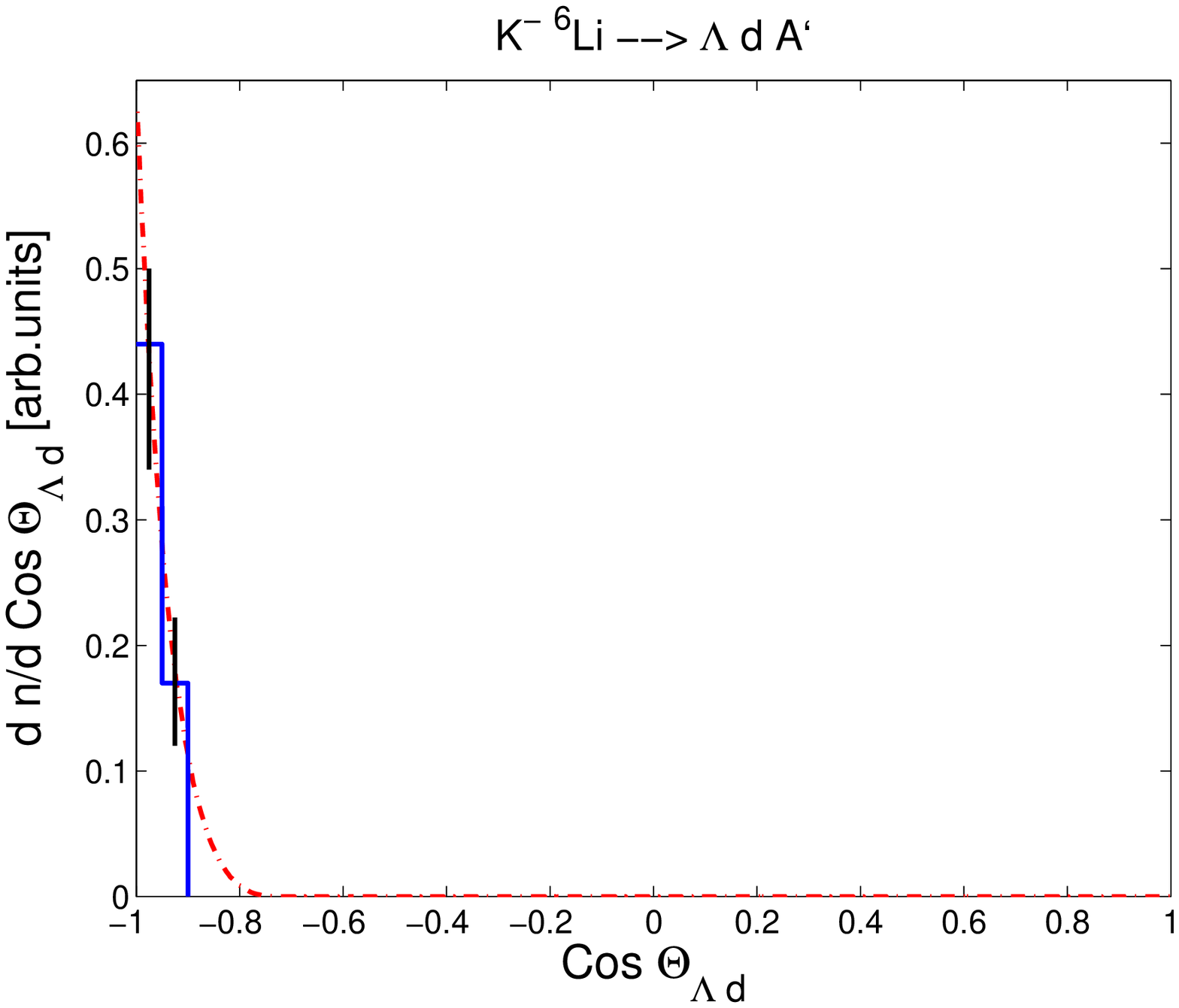}
\caption{The $\Lambda d$ invariant mass distribution (left plot) and $\Lambda d$ angular distribution (right plot) for $K^-$ absorption in $^{6}$Li. 
Histogram and error bars are from the experimental
paper \cite{:2007ph}, while the dot-dashed curve is the result of our calculation \cite{Magas:2008bp}.
}
\label{fig3}       % Give a unique label
\end{figure*}

		In the Refs. \cite{Magas:2008bp} the authors performed
       detailed calculations of $K^-$ absorption by three nucleons in $^6$Li 
       and show that all features observed in \cite{:2007ph} can be well
       interpreted in the picture of three body kaon absorption, as suggested in
    \cite{Suzuki:2007kn}, with the rest of the nucleons acting as
    spectators - see Fig. \ref{fig3}, and Fig. \ref{fig1}, left plot.
    
    It is also important to note that in the same FINUDA experiment  the $ ^{12}C$ was also 
    used as a target, and for it no clear peak has been observed in the  $\Lambda d$ invariant mass spectrum  \cite{:2007ph}. 
    This was attributed \cite{:2007ph} to final state interactions of the particles produced in the primary 
    absorption process with the residual nucleus, in complete agreement with the mechanisms discussed in Ref. \cite{Magas:2006fn}.

After having reported the latest achievements in the search of $K^-$ bound states with nucleons, let us draw some conclusions. 

First of all the peaks observed by FINUDA in the ($\Lambda p$)  \cite{Agnello:2005qj} and ($\Lambda d$) \cite{:2007ph} invariant mass distributions, after the absorption of stopped $K^-$ in different nuclei,  
are naturally explained in terms of: \\ 
\newpage
\noindent
- $K^-$ absorption by two \cite{Magas:2006fn,Kpp,Ramos:2007zz} or three \cite{Magas:2008bp} nucleons correspondingly, leaving the rest of the original nuclei as spectator; \\
- for the reactions on heavy nuclei the subsequent interactions of the particles produced in the primary absorption process ($\Lambda$, $p$, $d$ etc.) with the residual nucleus have to be taken into account. 

All the elements of the calculations in Refs. \cite{Magas:2006fn,Magas:2008bp}, namely \\
- proper kaon wave function, $\Psi_{K^-}(\vec{r})$,\\
- binding energy of the nucleons inside the nucleus,\\
- the momentum spread of the nucleons inside the nucleus due to Fermi motion,\\
- final state interactions after the primary absorption reaction,\\
are known; the calculations are straightforward and contain no free parameters. And from the simulations of the other reactions of hadrons with nuclei it is quite clear that all these elements have to be taken into account before we can make any conclusion about existence of bound $K^-$ states in nuclei. 

Such a modeling reproduces all the observables measured in the experiments \cite{Agnello:2005qj,:2007ph} and used to support the hypothesis of the deeply bound kaon state, see for example Figs. \ref{fig2}, \ref{fig3} and the left plot of Fig. \ref{fig1},   
therefore we can conclude that at present there is no experimental evidence of deeply bound $K^-$ state in nuclei. 

In the calculations of \cite{Magas:2008bp,Magas:2006fn} the positions and widths of the peaks, interpreted by the FINUDA collaboration as deeply bound kaonic states, do not depend on the exact strength of the $K^- N$ interaction and are completely determined by the kinematics of the studied process. Thus, we can conclude that the experimental study of $K^- N$ interaction at short distances is not yet possible, due to low accuracy of the data (for example the experimental $\Lambda p$ invariant mass spectrum is only available for the mixture of different nuclei) and also our limited knowledge of the $K^-$ absorption dynamics and final state interactions. 
Until these latter processes are not well understood we will not be able to make any progress in the search for possible $K^-$ deeply bound states in nuclei, which is aimed by future experimets, for example AMADEUS \cite{Amadeus}.

\section{What can we learn from the FINUDA data?}
\label{sec-2}

Although the FINUDA experiments have been done for reasons which are not supported a posteriori, some new physics,  related to kaon absorption dynamics and final state interactions of the particles produced in the primary absorption process,  can be extracted from the data. 

First of all, we have verified that $K^-$ absorption widths due
to two and three nucleons are consistent with a dependence on
the corresponding powers of the nuclear density:
\beq
\Gamma_{NN} \propto \int d^3 \vec{r}  |\Psi_{K^-}(r)|^2 \rho^2(r)\,, \quad \Gamma_{NNN} \propto \int d^3 \vec{r}  |\Psi_{K^-}(r)|^2 \rho^3(r)\,,
\eeq{e1}
where $\Psi_{K^-}(r)$ is the $K^-$ atomic wave function.

\begin{figure*}
%  \includegraphics[width=1.0\textwidth]{Gamma_3N_29_vs_3d.eps}
 \includegraphics[width=0.5\textwidth]{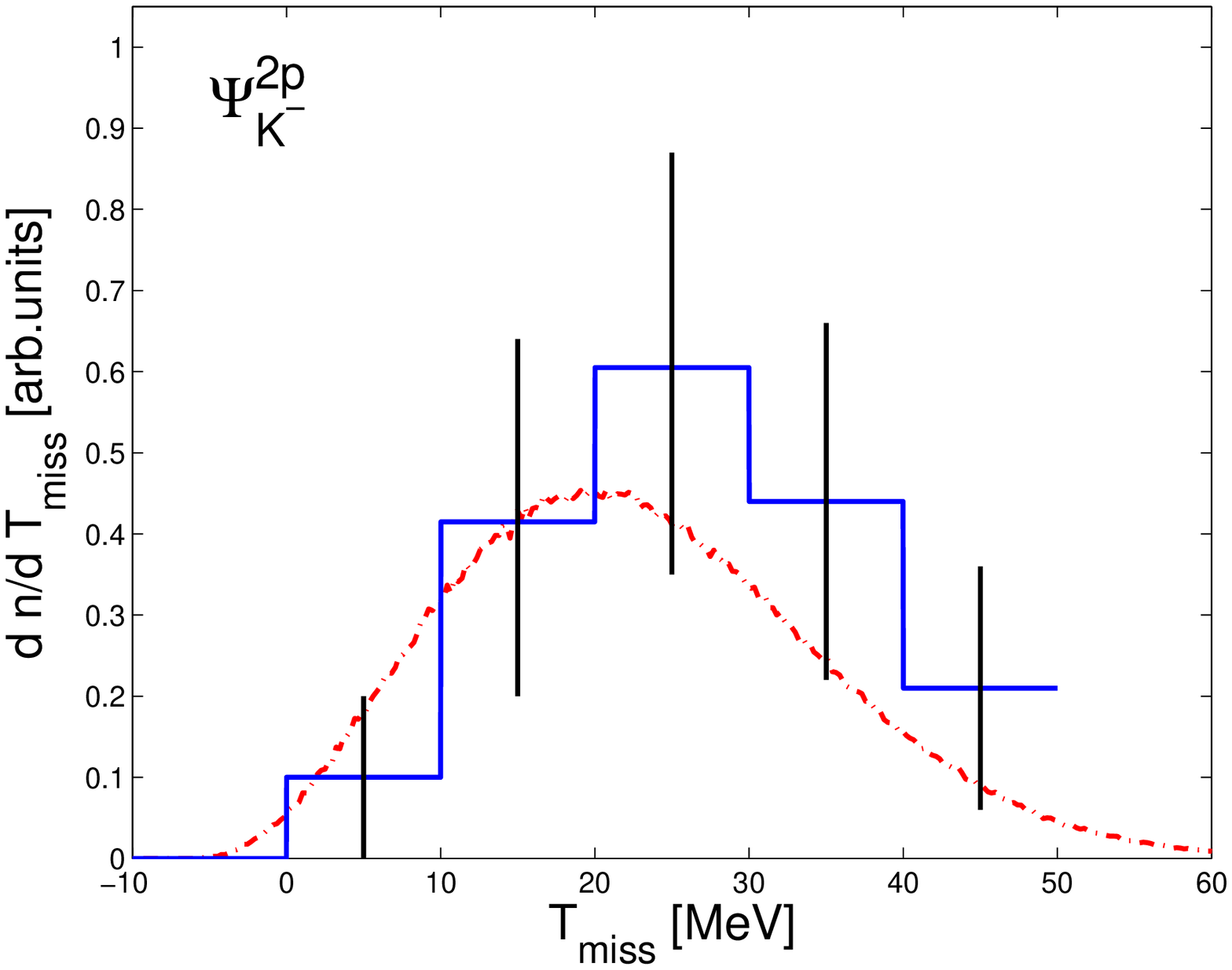}
 \includegraphics[width=0.5\textwidth]{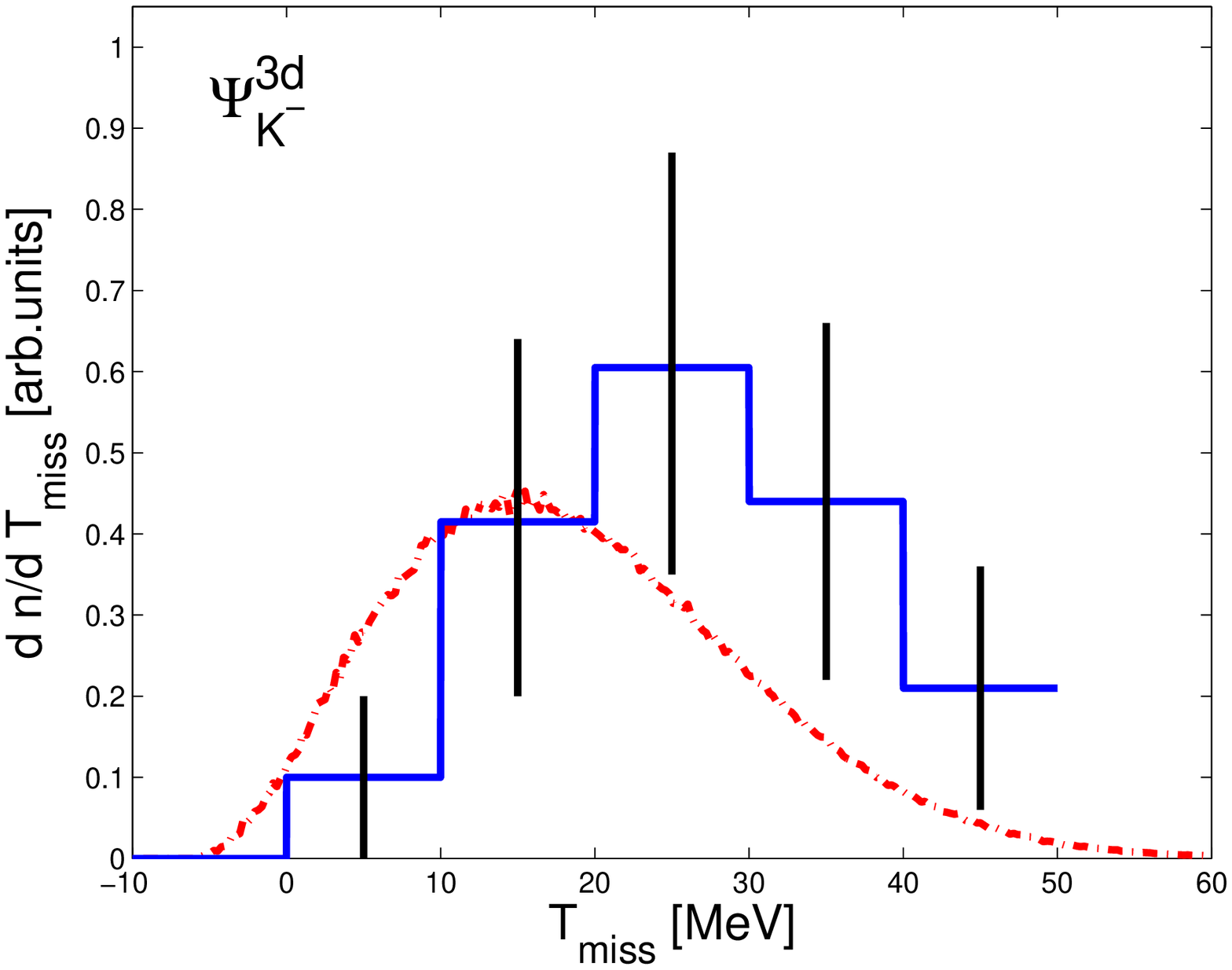}
\caption{The ($\Lambda d$)  missing mass distribution, $T_{miss} = m_{K^-}+M_{ ^{6}{\rm Li}} - m_\Lambda - m_n - 2 M_d - (T_\Lambda + T_d)$ \cite{:2007ph},  
for $K^-$ absorption at $^6$Li, calculated according to Ref. 
\cite{Magas:2008bp} using $2p$ (left plot) and $3d$ (right plot) $K^-$ wave functions, experimental histogram is from \cite{:2007ph}.  
The agreement for $\Psi_{K^-}^{2p}(r)$ is much better, as expected - see text for more details. }
\label{fig1}       % Give a unique label
\end{figure*}

Some information on $K^-$ wave function for different nuclei can also be extracted from the simulations \cite{Magas:2008bp,Magas:2006fn}. In Ref.  \cite{Magas:2006fn} the following way of choosing $\Psi_{K^-}(r)$ was suggested. The $K^-$ absorption at rest proceeds by capturing a slow $K^-$ in
a high atomic orbit of the nucleus, which later cascades down till
the $K^-$ reaches a low lying orbit, from where it is finally absorbed by the
nucleus  \cite{gal-klieb}.
Thus we assume the absorption to take
place from the lowest level  where the energy shift for atoms has been
measured, or, if it is not measured, from the level where the calculated
shift \cite{okumura} falls within the measurable range.
In the case of light nuclei $^6$Li, $^{7}$Li, $^{9}$Be, $^{12}$C, $^{13}$C, 
$^{16}$O the absorption takes place from the $2p$ orbit; for $^{27}$Al from $3d$ orbit; 
and for $^{51}$V from $4f$ orbit.  
To demonstrate the sensitivity of our simulations to the choice of $\Psi_{K^-}(r)$ we show on Fig. \ref{fig1} 
the ($\Lambda d$) missing mass distribution for $K^-$ absorption at $^6$Li calculated according Ref. 
\cite{Magas:2008bp} using $2p$ and $3d$  $K^-$ wave functions. Clearly the agreement for $\Psi_{K^-}^{2p}(r)$ is 
much better.

The nucleon and $\Lambda$ from the primary  absorption reactions  re-scatter with
nucleons in the remnant nucleus according to a probability per unit length
given by $\sigma \rho(r)$, where $\sigma$ is the
experimental $NN$ or $\Lambda N$ cross section at the corresponding energy. 
In \cite{Magas:2006fn} a simpler parameterization for the $\Lambda$ cross section  was employed: 
$\sigma_\Lambda=2\sigma_N/3$.  
In \cite{Kpp} we improved on this taking the parameterization of $\sigma_{\Lambda}$ cross 
section from Ref. \cite{manolo}:
\begin{equation}
\sigma_{\Lambda} =(39.66-100.45 x+92.44 x^2-21.40 x^3)/p_{LAB}\ {\rm [mb]}\,,
\label{sLN}
\end{equation}
where $x=Min(2.1$ GeV, $p_{LAB})$. 
In Ref. \cite{Kpp} the ($\Lambda p$) invariant mass distributions were calculated  for $^6$Li, $^7$Li, $^{12}$C, $^{27}$Al and $^{51}$V (the targets
of the FINUDA experiment \cite{Agnello:2005qj}) and $^9$Be, $^{13}$C, $^{16}$O (to be included into the future FINUDA experiment \cite{future_finuda}). 
The simulations show that the modification of $\sigma_{\Lambda}$ affects non-negligibly only the results for heavy nuclei.  However, 
to verify our model of $K^-$ two nucleon absorption dynamics, our choices of $\Psi_{K^-}(r)$, and our simulations of final state interactions 
experimental spectra for separate nuclei are necessary, while at the moment the ($\Lambda p$) invariant mass distribution is only available for the mixture of three lightest targets \cite{Agnello:2005qj}.

%\begin{acknowledgements}
%\end{acknowledgements}



\begin{acknowledgements}
The authors thank H. Toki for fruitful and enlightening discussions. 
This work is partly supported by contracts FIS2006-03438 and
FIS2005-03142 from MEC (Spain) and FEDER, the Generalitat de
Catalunya contract 2005SGR-00343 and by the Generalitat Valenciana. 
This research is part of the EU
Integrated Infrastructure Initiative Hadron Physics Project under
contract number RII3-CT-2004-506078, and V.K.M.
wishes to acknowledge direct support from it. 
\end{acknowledgements}

\end{document}